\documentstyle[twocolumn,aps,epsf]{revtex}

\newcommand{\Qv}{\mbox{{\bf Q}}}

\newcommand{\Av}{\mbox{{\bf A}}}

\newcommand{\Rv}{\mbox{{\bf R}}}

\newcommand{\lv}{\mbox{{\bf l}}}

\begin{document}
\draft
\title{Upper critical field for underdoped high--$T_c$ superconductors. Pseudogap and
stripe--phase.}
\author{Marcin Mierzejewski and Maciej M. Ma\'ska\cite{email}}
\address{
Department of Theoretical Physics, Institute of Physics, 
University of Silesia, 40-007 Katowice,
Poland}
\maketitle
\begin{abstract}

We investigate the upper critical field in a
stripe--phase and in the presence of a phenomenological pseudogap.
Our results indicate that the formation of stripes affects the Landau 
orbits and results in an enhancement of $H_{c2}$. On the other hand,
phenomenologically introduced pseudogap leads to a reduction of the 
upper critical field. This effect is of particular importance when 
the magnitude of the gap is of the order of the superconducting 
transition temperature. We have found that a suppression of the
upper critical field takes place also for the gap that originates
from the charge--density waves.

\end{abstract}
\section{Introduction}

The high--temperature superconductors (HTSC) exhibit qualitative
differences with respect to the classical
superconducting systems. Normal state properties 
of underdoped superconductors and the upward curvature of the
upper critical field ($H_{c2}$) belong to one of the most spectacular 
examples. The presence of a normal--state pseudogap has been
confirmed with the help of different experimental techniques like:
angle--resolved photoemission \cite{arpes1,arpes2,arpes3},
intrinsic tunneling spectroscopy \cite{krasnov1,krasnov2}, 
NMR \cite{nmr1,nmr2},
infrared \cite{infra} and transport \cite{transport} measurements.
Despite a wide spectrum of experimental data the underlying
microscopic mechanism if far from being understood.
A tempting hypothesis that the pseudogap is a precursor of the 
superconducting gap has not definitively been confirmed.
In particular, the neutron scattering experiments 
\cite{muller} reveal qualitative differences between the
isotope effects observed for the superconductivity
and the pseudogap. Moreover, results obtained with the help of 
intrinsic tunneling spectroscopy  \cite{krasnov1,krasnov2}
speak against the superconducting origin of the pseudogap. 
The coexistence of superconductivity and the charge--density waves
\cite{eremin,seibold,markiewicz,kramer}
can be considered as a possible scenario, that accounts
for differences between superconductivity and the normal--state
gap. Inhomogeneous distribution of holes which enter the 
copper-oxygen planes in the doping process can give rise to the
formation of stripes. There is a convincing experimental 
and theoretical argumentation speaking in favor of the
stripe--phase \cite{str1,str2,str3,str4,str5,str6,str7}
with an intimate connection between
superconducting and stripe correlations. 
This phase consists of antiferromagnetic domains
which are separated by the hole--rich domain walls.

Differences between the high--temperature 
superconductors and classical systems show also up in the
magnetic properties. The high--$T_c$ compounds are
characterized by large values of the upper critical field
and its unusual temperature dependence. The resistivity
measurements clearly indicate an upward
curvature of the upper critical field with no evidence
of saturation even at genuinely low temperatures 
\cite{osofsky,mackenzie}. These results remain
in disagreement with the conventional, microscopic
approach \cite{gorkov}. This discrepancy can be
explained as a result of the Josephson tunneling between
superconducting clusters \cite{gesh,wen} produced by a
macroscopic phase separation.

Due to the complexity of the Gor'kov equations one usually assumes
that the normal--state properties of the system under consideration
can properly be described by three-- \cite{hw} or two--dimensional
\cite{kresin} electron gas. Recently, we have proposed an approach
that enables calculation of the upper critical field for
a two--dimensional lattice gas \cite{my1,my2}. This method 
allows one to derive $H_{c2}$ in a similar way as one
calculates the 
critical temperature  in the standard
BCS formalism. Therefore, any extension of the analysis of 
the upper critical field is rather straightforward.
In the present paper we calculate the upper critical field
in a system that exhibits some important properties
of hole--doped cuprates: stripe--phase 
and the presence of the pseudogap. 
In the latter case we discuss
the coexistence of superconductivity and charge--density--wave 
as well as a phenomenological pseudogap. We show that
the anomalous properties of the high--temperature
superconductors are reflected in 
the upper critical field.  

\section{$H_{c2}$ in the presence of charge--density--waves}

We consider a two--dimensional square lattice immersed
in a perpendicular uniform magnetic field of magnitude $H_z$. 
We assume
the nearest--neighbor pairing interaction, $H_V$, that is responsible
for anisotropic superconductivity and the interaction term,
$H_{\rm CDW}$,
which leads to the charge--density--waves. The relevant
Hamiltonian reads \cite{my1,my2}
\begin{equation}
H=H_0+H_V+H_{\rm CDW},
\end{equation}
where
\begin{eqnarray}
H_0&=&\sum_{\langle ij\rangle,\sigma} t_{ij}\left(\Av \right) 
c^{\dagger}_{i\sigma}c_{j\sigma} \nonumber \\
&&
 +g\mu_{B}H_{z}\sum_{i} 
\left(c^{\dagger}_{i\uparrow}c_{i\uparrow}
-c^{\dagger}_{i\downarrow}c_{i\downarrow} \right).
\end{eqnarray}
Here, $c^{\left(\dagger\right)}_{i\sigma}$ annihilates
(creates) an electron with spin $\sigma$ at the lattice
site $i$, 
$g$ is the gyromagnetic ratio and $\mu_B$ is the Bohr
magneton. 
$t_{ij}\left(\Av \right)$ is the nearest--neighbor
hopping integral that in the presence of the magnetic field
acquires the Peierls phase--factor \cite{peir,my1}
\begin{equation}
t_{ij}\left(\Av \right)= t\,
\exp\left(\frac{ie}{\hbar c} \int^{\Rv_{i}}_{\Rv_{j}}
\Av\cdot d\lv\right).
\end{equation}
In the mean-field approximation the pairing 
interaction and the CDW coupling take on the form
\begin{eqnarray}
H_{V}&=&-\:
V\sum_{\langle ij\rangle}\left(c^{\dagger}_{i\uparrow}c^{\dagger}_{j\downarrow}
\Delta_{ij}
+c_{i\downarrow}c_{j\uparrow} \Delta^{*}_{ij} \right), \\
H_{\rm CDW}&=&-\delta_{\rm CDW} \sum_{j \sigma} 
e^{i \Qv \cdot \Rv_{j}} 
c^{\dagger}_{j \sigma} c_{j \sigma},  
\end{eqnarray}
where $\Delta_{ij}=\langle c_{i\downarrow}c_{j\uparrow}
-c_{i\uparrow}c_{j\downarrow}\rangle$ is the superconducting singlet order
parameter and $\delta_{\rm CDW}$ represents the magnitude of the 
CDW gap.
The complexity of calculations strongly depends on the
CDW modulation vector $\Qv$.
For the sake of simplicity we consider a commensurate
charge--density wave with 
$\Qv=\left(\pi,\pi\right)$. 
This choice of the modulation vector results in the
gap in the density of states that opens in the middle of the band
(in our case at the Fermi level) independently on the magnitude
of the external magnetic field.
Since the pseudogap hardly depends on the magnitude of the
magnetic field \cite{krasnov2}, $\delta_{\rm CDW}$ will be taken as 
a model parameter.

In order
to calculate the upper critical field we make use of the unitary
transformation that diagonalizes the kinetic part of the
Hamiltonian \cite{my1,my2}. In the case of the Landau gauge, 
$\Av=H_z\left(0,x,0\right)$, 
this transformation is determined by a plane--wave function
in $y$ direction  
and an eigenfunction of the Harper equation
\cite{harper}:
\begin{eqnarray}
&& g\left(\bar{k},p,m+1\right)+
2\cos\left(h m -pa\right)g\left(\bar{k},p,m\right)
\nonumber \\
&&+ g\left(\bar{k},p,m-1\right)=
t^{-1}E\left(\bar{k},p\right)
g\left(\bar{k},p,m\right).
\end{eqnarray}
Here, $m$ is an integer number that enumerates
the lattice sites in $x$ direction,
whereas $h$ is the reduced magnetic field, $h= 2 \pi \Phi/\Phi_0$, 
that is expressed by the ratio of the flux $\Phi$
through the lattice cell and the flux quantum $\Phi_0$.
$p$ is the wave--vector in $y$ direction and $\bar{k}$
is an additional quantum number, that in the absence
of the magnetic field is the wave--vector in
$x$ direction.
In the new basis the normal--state Hamiltonian takes on the form:
\begin{eqnarray}
H_{0}  &=& \sum_{\bar{k},p,\sigma} E_{\bar{k}p\sigma}
a^{\dagger}_{\bar{k}p\sigma}a_{\bar{k}p\sigma}, \\
H_{\rm CDW}&=& -\delta_{\rm CDW}\!\! \sum_{\bar{k},\bar{l},p,m,\sigma}
g^*\left(\bar{k},p+\pi,m\right)g\left(\bar{l},p,m\right) \nonumber \\
&& \times\: e^{i \pi m } 
a^{\dagger}_{\bar{k}p+\pi,\sigma}a_{\bar{l}p\sigma},
\end{eqnarray}
where
\begin{equation}
E_{\bar{k}p\sigma}=E\left(\bar{k},p\right)+\sigma
g\mu_{B}H_{z}. 
\end{equation}

One can prove that if $E$ represents the eigenvalue of
the Harper equation obtained for the wave--vector $p$ then
$-E$ is one of the eigenvalues corresponding to $p+\pi$.
It can also be shown that  
$ \tilde g\left(\bar{k},p,m\right)= 
g\left(\bar{k},p+\pi,m\right)\exp\left(i \pi m \right)$
represents an eigenfunction of the Harper equation
calculated for momentum $p$   
with the eigenvalue $-E\left(\bar{k},p+\pi\right)$.
With the help of these relations one can obtain analytically
the energy spectrum of the Hamiltonian in the normal state,
$H_0+H_{\rm CDW}$, provided that the eigenvalues of the Harper
equation are known. In particular, one can calculate the 
anomalous Green functions which are related to the 
superconducting order parameter:
\begin{eqnarray}
&& \langle\langle a_{\bar{l}p\uparrow}\mid a_{\bar{k},-p\downarrow}\rangle\rangle = 
\nonumber \\
&& -\sum_{m}\left[ X^{*}_{\bar{l}\bar{k}p}\left(m\right) 
\Delta^{x}\left(m\right)+Y^{*}_{\bar{l}\bar{k}p}\left(m\right) 
\Delta^{y}\left(m\right)\right] K_{\bar{l}\bar{k}p}(\omega), \nonumber \\
\end{eqnarray}
where $X$ and $Y$ are determined by the solution of the Harper equation
\begin{eqnarray}
X_{\bar{l}\bar{k}p}\left(m\right) &=& 
g\left(\bar{l},p,m\right)g\left(\bar{k},-p,m+1\right) \nonumber \\ 
&& + g\left(\bar{l},p,m+1\right)g\left(\bar{k},-p,m\right), \\ 
 Y_{\bar{l}\bar{k}p}\left(m\right) &=& 2 \cos(p) 
g\left(\bar{l},p,m\right) g\left(\bar{k},-p,m\right).
\end{eqnarray}
Due to the plane--wave behavior in $y$ direction the superconducting
order parameter depends explicitly only on the position in $x$ direction
(see Ref. \cite{my2} for the details)
\begin{eqnarray}
\Delta^{x}\left(m\right) &=& 
\frac{V}{N} \sum_{\bar{l}\bar{k} p}
X_{\bar{l}\bar{k}p}\left(m\right) \langle
a_{\bar{k},-p\downarrow}a_{\bar{l}p\uparrow}
\rangle, 
\\
\Delta^{y}\left(m\right) &=& 
\frac{V}{N} \sum_{\bar{l}\bar{k} p}
Y_{\bar{l}\bar{k}p}\left(m\right) \langle
a_{\bar{k},-p\downarrow}a_{\bar{l}p\uparrow}
\rangle.
\end{eqnarray}
$K_{\bar{l}\bar{k}p}(\omega)$, when 
integrated over $\omega$ with the Fermi function, gives the Cooper pair
susceptibility
\begin{equation}
K_{\bar{l}\bar{k}p}(\omega)=
\frac{
\left(\omega+E_{\bar{l}p\uparrow}\right)\left(\omega-E_{\bar{k},-p\downarrow}\right)
+\delta_{\rm CDW}^2
}{
\left(\omega^2-E^2_{\bar{l}p\uparrow}-\delta_{\rm CDW}^2\right)
\left(\omega^2-E^2_{\bar{k},-p\downarrow}-\delta_{\rm CDW}^2\right)
}.
\end{equation}
One can see that the impact of the charge--density waves on superconductivity
is brought about only by the modification of this quantity. Equations (10),
(13) and (14) allow one to calculate the upper critical field. It is determined
as the highest magnitude of the magnetic field for which there exists
a non--zero solution for $\Delta^{x}(m)$ and $\Delta^{y}(m)$.

\begin{figure}
\epsfxsize=9cm
\centerline{\epsffile{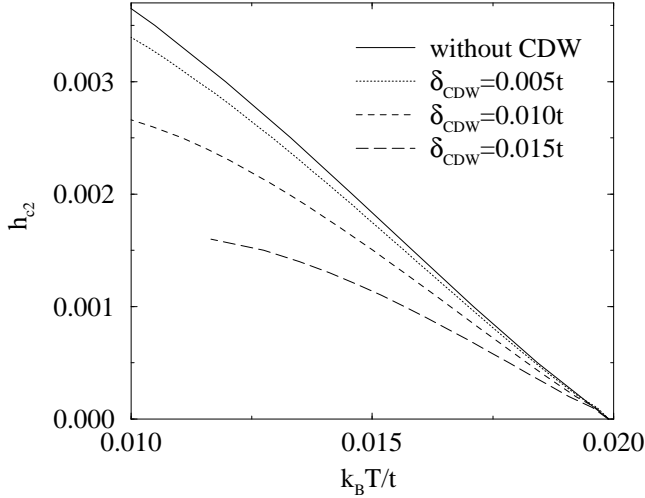}}
\caption{Temperature dependence of the upper critical field for different
magnitudes of the CDW gap.}
\end{figure}

Figure 1. shows the reduced upper critical field, 
$h_{c2}$, as a function of temperature calculated for different magnitudes 
of the CDW order parameter, $\delta_{\rm CDW}$. We have adjusted the magnitude 
of the pairing potential $V$ that gives the same superconducting 
transition temperature $kT_{c}=0.02t$ for all values of $\delta_{\rm CDW}$.  
These results have been obtained
for $120 \times 120$ cluster that at temperatures $kT \sim 10^{-2} t$ 
gives convergent results (we refer to Ref. \cite{my2} for details
of the cluster calculations).
One can see that even for small magnitudes of the CDW order parameter
the upper critical field is significantly reduced. However, qualitative
temperature dependence of $H_{c2}$ 
is not affected by the charge--density wave correlations. 
The reduction of the upper critical field due to the charge--density waves
may be brought about by a direct coupling between CDW and superconducting order 
parameters as well as by the modification of the density of states. In order to 
distinguish these contributions we investigate the upper critical field
in the presence of a phenomenological normal--state gap of arbitrary
magnitude and depth. This problem will be discussed in the next section. 
%%%

\section{$H_{c2}$ in the presence of a phenomenological gap}

In this section we investigate modification of the upper critical
field that originates only from the normal--state gap in the density of states.
In contradistinction to the analysis presented
in the previous section, the density of states may remain finite despite 
the presence of the gap. Here, the normal--state gap is characterized by
the width $2\delta$ and the relative depth 
$\left(\rho_0-\rho_{\rm PG}\right)/\rho_0$, where $\rho_{\rm PG}$ and
$\rho_0$ denote the density of states in the presence and without the
pseudogap, respectively. It is visualized in the inset in Fig. 2.
In the absence of the CDW order the upper critical field is determined
by Eq. (10) with
\begin{equation}
K_{\bar{l}\bar{k}p}(\omega)=
\frac{
1
}{
\left(\omega-E_{\bar{l}p\uparrow}\right)
\left(\omega+E_{\bar{k},-p\downarrow}\right)
}.\end{equation}
In order to account for the modification of the density of states
we renormalize the normal--state propagators which give rise
to the Cooper-pair susceptibility
\begin{eqnarray}
&& \frac{1}{\omega-E_{\bar{l}p\uparrow}} \longrightarrow 
\frac{\rho_{\rm PG}}{\rho_0} \frac{1}{\omega-E_{\bar{l}p\uparrow}} 
+\frac{\rho_0-\rho_{\rm PG}}{2\rho_0} \nonumber \\
&& \times 
\left[ \left(1+\frac{E_{\bar{l}p\uparrow}}{\sqrt{E^2_{\bar{l}p\uparrow}+\delta^2}} 
\right)\frac{1}{\omega-\sqrt{E^2_{\bar{l}p\uparrow}+\delta^2}} \right.  
\nonumber \\
&&+\left. 
\left(1-\frac{E_{\bar{l}p\uparrow}}{\sqrt{E^2_{\bar{l}p\uparrow}+\delta^2}} 
\right)\frac{1}{\omega+\sqrt{E^2_{\bar{l}p\uparrow}+\delta^2}}
\right] \nonumber. \\
\end{eqnarray} 
In the limiting case $\rho_{\rm PG}=\rho_0$ one obtains the standard density of states
as determined by the Hofstadter spectrum, whereas  
for $\rho_{\rm PG}=0$ the density of states vanishes in the vicinity 
of the Fermi energy. Substituting the renormalized
propagators into Eq. (16) one can calculate the upper critical 
field in the same way as described in the previous section.

\begin{figure}
\epsfxsize=9cm
\centerline{\epsffile{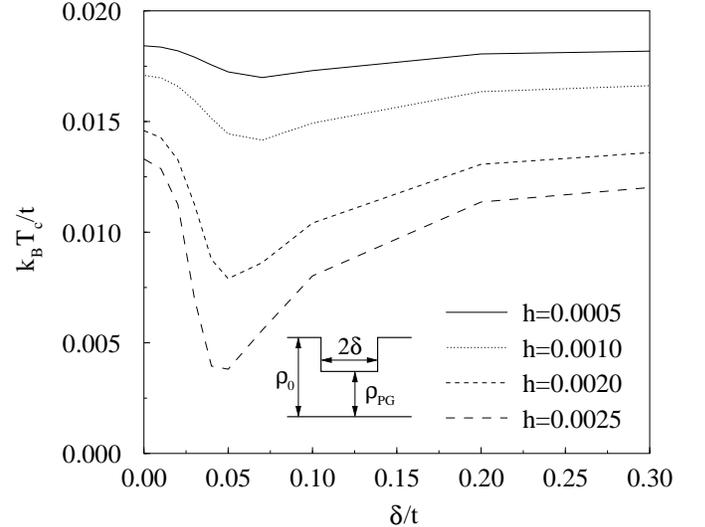}}
\caption{Critical temperature as a function of the pseudogap width $\delta$
for different values of the external magnetic field. $\rho_{\rm PG} = 
\frac{1}{2} \rho_0$ was used. The inset shows a schematic
density of states in the vicinity of the Fermi level.}
\end{figure}

Figure 2. shows the superconducting transition temperature
obtained for different values of the reduced magnetic field 
with $\rho_{\rm PG}=\frac{1}{2} \rho_0$.  As before, the intersite coupling 
$V$ have been adjusted to obtain $kT_c=0.02t$ in the absence of magnetic field.
One can see that the upper critical
field is reduced due to the presence of the normal--state gap. 
The most significant lowering of
$H_{c2}$ takes place for finite values of the $\delta$
which are comparable to the magnitude 
of the superconducting gap. This result originates from the fact
that the Cooper pair susceptibility is strongly peaked at the Fermi level with
a characteristic energy scale that is determined by temperature.
Therefore, for $\delta \gg kT_{c}$ the pseudogap results in a global lowering
of the density of states which can be compensated by an enhancement of the
pairing potential. It means that assuming stronger pairing potential $V$
we can reproduce $H_{c2}(T)$ calculated in the absence of the gap.

\begin{figure}
\epsfxsize=9cm
\centerline{\epsffile{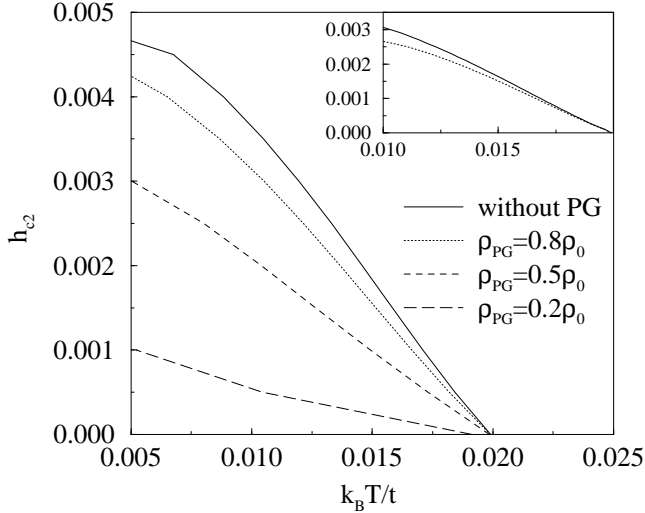}}
\caption{Upper critical field as a function of temperature for different
densities of states at the Fermi level. The half width of the pseudogap
is $\delta=0.1t$. The inset shows temperature dependence of upper critical 
field calculated for the phenomenological pseudogap with $\rho_{\rm PG}=0$
(solid line) and for the CDW gap (dashed line). In both cases the half 
width is $\delta=\delta_{\rm CDW}=0.01t$.}
\end{figure}

We have found that
the reduction of the upper critical field increases
with the depth of the gap as depicted in Fig. 3. 
The inset in Fig. 3 shows a comparison of $H_{c2}$ obtained for
the phenomenological pseudogap with $\rho_{\rm PG}=0$ and for the
charge--density waves. A comparison of these results clearly
indicate that the coupling between CDW and superconducting 
order parameters results in a small decrement of the
upper critical field.

\section{$H_{c2}$ in a stripe phase}
Other unusual feature of HTSC, that we discuss in the present section,
is related to inhomogeneous distribution of holes.
It results in a stripe--phase which consists of antiferromagnetic 
domains separated by hole--rich domain walls.
We study how the upper critical field is affected by this specific 
distribution of carriers. 
In order to simulate the presence 
of a stripe--phase we carry out the calculations for a long and narrow 
rectangular--shape clusters. 
We assume that the isolating, antiferromagnetic domains can
be simulated by fixed boundary conditions in the direction
perpendicular to the stripes (along the $x$ axis). 
The spatial organization of the stripe structure has been 
intensively investigated on experimental \cite{trank,stripee} and
theoretical grounds \cite{scalap1,scalap2}.
Experimental data for HTSC show that
the width of stripes depends on the concentration of holes and is
of the order of a few lattice constants.
The neutron--scattering study of the stripe phase \cite{trank}
suggests that the hole-rich domain walls are only single cell wide. 
On the other hand, the numerical study of the two--dimensional $t$--$J$ model
\cite{scalap1} shows that the domain walls may have a significant
density of holes over three rows of sites.
According to these results we consider $150 \times n$
finite systems, where $n=2,3$ and $7$.
Our simplified approach does not restore the
actual structure of the stripe--phase. In particular, for $n=1$ one
obtains an unphysical, purely one--dimensional system, that hardly depends
on the external magnetic field. Therefore, we investigate the
rectangular--shape clusters with the width as a free parameter.
Since we neglect the correlations between different stripes, 
the upper critical field is determined by Eqs. (10) and (16).

In the case of free electron gas external magnetic field leads 
to the occurrence of rotationally invariant states corresponding 
to the Landau orbits. 
However, the geometry of the stripe--phase may seriously
affect the formation of the Landau orbits. This effect is of particular
significance if the radii of the Landau orbits, $R_L$, exceed the width 
of the stripe, $an$, ($a$ is the lattice constant). 
In order to visualize the impact of magnetic field on electrons in 
the stripe--phase we have calculated the resulting current distribution.
Within the framework of the linear--response theory the current operator 
is given by $\hat{J}_l(x,y)=-\partial \hat{H}/\partial A_l(x,y)$, 
where $(x,y)$ denotes spatial coordinates and $l$'s are unit 
vectors in the lattice axes directions. Results obtained
in the normal state ($V=0$)
on a $150 \times 7$ cluster with applied magnetic field 
$h=0.1$  are presented in Figure 4. 

\begin{figure}
\epsfxsize=10cm
\centerline{\epsffile{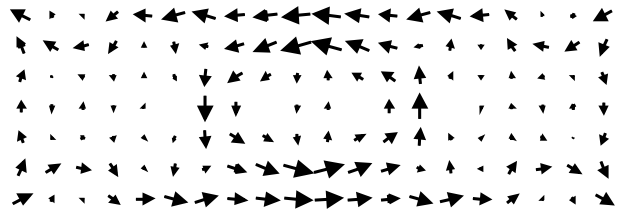}}
\caption{Current distribution in a piece of stripe of width of 7 sites, 
calculated for the reduced magnetic field $h=0.1$. The lengths of the arrows
are proportional to the currents. Such a pattern is periodically repeated
along the stripe.}
\end{figure}

Modification of the Landau orbits affects 
the diamagnetic pair--breaking mechanism. Therefore, 
one may expect that superconductivity survives in
the presence of much stronger magnetic fields than in the
homogeneous phase. This observation is confirmed by the numerical 
calculations, as depicted in Figure 5. 
\begin{figure}
\epsfxsize=9cm
\centerline{\epsffile{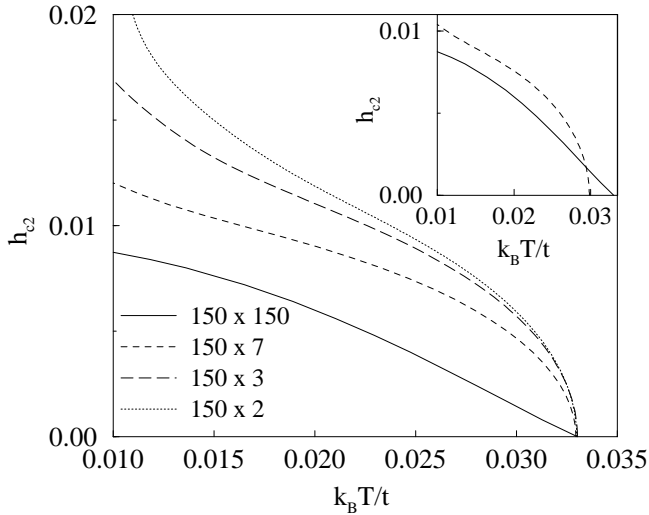}}
\caption{Upper critical field as a function of temperature calculated 
for stripes of different width. We have chosen appropriate values of $V$
which give the same transition temperature in the absence of magnetic field.
The inset shows a comparison of upper critical field for $150\times 150$
and $150\times 7$ systems with the same value of the pairing strength 
$V=0.244t$
}
\end{figure}

\noindent
Here, $150 \times 150 $ 
cluster corresponds to an infinite system. The enhancement of $H_{c2}$ 
is of particular 
importance for weak magnetic fields, when $R_L/na \rightarrow \infty$.
One can observe a dramatic change of the 
slope, ${\rm d}H_{c2}/{\rm d}T$, calculated at $T=T_c$. Here,
the impact of the magnetic field on the superconducting transition 
temperature is much less than in the homogeneous two--dimensional case.

The pseudogap and stripes affect the superconducting
properties of the system both in the presence and in the absence of
the magnetic field. Modification of the density of states changes the
effective coupling constant, $\lambda=\rho_{\rm FS} V$, that enters the 
standard BCS gap equation.  
Therefore, we have directly compared the $H_{c2}$ for systems, which in the
absence of magnetic field are characterized by the same transition
temperature (one can roughly say that $\lambda={\rm const}$). 
In order to complete the discussion, we have also calculated
the $H_{c2}(T)$ for the case when the pairing potential does not 
depend on the pseudogap and the stripe structure ($V={\rm const}$). Since, the 
opening
of the pseudogap reduces $T_c$ it results also in an additional decrement 
of the upper critical field, when compared to the results presented in Figs. (1-3). 
However, an enhancement of the $H_{c2}$ in the stripe phase can take place 
despite
the reduction of the superconducting transition temperature, as depicted in the
inset in Fig. 5.

\section{Concluding remarks}

In order to clarify some physical aspects of our method one can
compare it with approaches, which are commonly used 
to investigate $H_{c2}$. Previously, we have applied
the same method to discuss the upper critical
field for isotropic superconductivity \cite{my1,my2}.
Then, one ends up with the gap equation that can be written
in the form  
\begin{equation}
\Delta_{i}= \frac{V}{\beta }\sum_{j,\omega_n}
\Delta_{j} G(i,j,\omega_n)G(i,j,-\omega_n). 
\end{equation} 
Here, $\Delta_i=\left<c_{i\downarrow} c_{i \uparrow} \right>$ and
$G(i,j,\omega_n)$ is the one--electron Green's function
in the presence of a uniform and static magnetic field.
It is clear that the above equation is a
lattice version of the linearized Gor'kov equations 
\cite{gorkov}, which determine
the critical field at a second-order transition,
where the superconducting gap $\Delta_i$  is 
vanishing \cite{hw1964}. However, our method 
does not allow to discuss the superconducting properties
below the $H_{c2}$ (e.g. the vortex state).  In our approach
the electron Green's functions have been calculated exactly,
whereas in the standard case one makes use of
the semiclassical approximation that neglects the Landau
level quantization.   

To conclude, 
we have investigated how the upper critical field is connected with different
features of high--temperature superconductors. In particular, we have 
discussed $H_{c2}$ in the presence of charge--density waves, phenomenological 
pseudogap and stripes. Our results suggest that a gap in the density of states 
reduces the upper critical field, independently on the underlying
microscopic mechanism. For finite density of states at the Fermi level
this reduction is mostly pronounced when the width of the gap is of the
order of the superconducting transition temperature. In the phase with 
isotropic CDW gap the density of states at the Fermi level vanishes.
Then, as one can expect, the upper critical field is strongly reduced 
even by a relatively small gap. Here, the coupling between the CDW
and superconducting order parameters results in an additional reduction 
of $H_{c2}$. On the other hand, in the presence of stripes the upper 
critical field is enhanced, especially close to $T_c$. We attribute 
this effect to the reduction of the orbital pair--breaking 
mechanism since the radii of the Landau orbits are much larger than
the width of the stripes.

The presented investigation of $H_{c2}$
is restricted to the simplest case of the uniform magnetic field and
neglects a possible disorder in the vortex system.
It can originate from fluctuations close to the phase transition
 or inhomogeneous charge and spin
distribution in the stripe--phase.
However, as we are concerned exclusively with the critical field
at the second--order transition, these effects are of minor importance.
We have also not discussed 
the reentrance of superconductivity in the strong--magnetic field.
This effect has been investigated in the continuum
model \cite{tesanovic,macdonal} as well as in the case of lattice
gas \cite{my1}, when the structure of fractal energy spectrum is reflected
in phase diagram. Theoretical argumentation that supports the reentrance of
superconductivity remains valid  also in the presence of pseudogap, at least
on the simplest level that has been used in the present paper.  
However, in the genuinely strong magnetic field the
assumption of the field--independent gap is unphysical and microscopic 
investigation of this phenomenon is needed.

\acknowledgments
This work has been supported by the Polish State Committee
for Scientific Research, Grant No. 2 P03B 01819. 
We acknowledge a fruitful discussion with Janusz Zieli{\'n}ski.

\end{document}